\documentstyle[12pt,epsfig]{article}
\title{On the distinct observation of the moving triangle singularity}
\author{V.M.Kolybasov \thanks{e--mail: kolybasv@sci.lebedev.ru}
 \\ Lebedev Physical Institute, 117924 Moscow, Russia}
\date{}
\begin{document}
\maketitle
\begin{abstract}
\normalsize
\baselineskip=24pt
It is shown that nonzero orbital momentum in the vertex of secondary
interaction in the triangle graph leads to a more clear picture
corresponding to the moving complex singularity compared with
the case of constant vertex. The peak in the
distribution of the final particles invariant mass becomes narrower
and its shift with momentum transfer more distinct. It opens the
possibility to observe this picture experimentally.
\\[\baselineskip]
\end{abstract}

\newpage
\baselineskip=24pt

The triangle mechanism (see Fig.1 graph) is one of the simplest
mechanisms of direct nuclear interactions. It is very frequently used
in the description of nuclear reactions. A characteristic graph of this
type corresponds to a primary interaction (the right upper vertex) followed
by secondary elastic or inelastic interaction (the lower vertex). The
experimental determination of its contribution as well as separation
of the kinematical regions of its domination would permit us to extract
valuable and reliable information on nuclear structure and the
off--shell amplitude of the ``elementary'' process in the lower vertex.

Until now very few cases are known when the dominant contribution of the
triangle mechanism is seen. Let us point to the peak in pion spectrum
from the capture of stopped kaons on the deuteron ${\mbox{K}}^- d \to
\mbox{p} \Lambda {\pi}^-$ [1] due to the triangle graph with N$\Sigma$
intermediate state and the conversion $\mbox{N} \Sigma \to \mbox{p} \Lambda$
[2]. Besides it is possible to mention anomalies in photonuclear
interactions [3] and enhancements in two nucleon mass spectra near
$2 {\mbox{m}}_{\mbox{N}}$ due to FSI [4]. On the other side, it is well
known that the characteristic feature of the triangle graph in nuclear
physics consists in so called moving complex triangle singularity. Its
position in the invariant mass $M$ of the lower group of final particles
depends on the value $q$ of three-momentum transfer
from initial to final fast particles in the right vertex [5]. The
distribution with respect to $M$ depends on the value of $q$. It is a
model--independent criterion of the dominant role of the graph [6].
Up to now this test was not used for the identification of the triangle
mechanism. The estimations for the process $pd \to pd{\eta}$ [7] and for
several other processes with an intermediate $\Delta$--isobar [8] predict
a clear picture of the moving singularity but there will be rather broad
maxima in the $M$ distributions with widths of the order of
$100 \div 200$ MeV.

Of course, the most distinct picture could be expected in the region of
the excitation energy $E_{\mbox{ex}}=M-(m_1 + m_2)$ of the order of
nuclear binding energies and momentum transfer $q$ of the order of nuclear
Fermi--momenta. Here the complex triangle singularity comes close to the
physical region and narrow peaks and background suppression could be
anticipated. Moreover, the cross sections in this region are much larger.
It is hindered by the fact that Fig.1 triangle graph has also the
threshold root singularity at $E_{\mbox{ex}}=0$ and usually it determines
a picture of the spectrum at small $E_{\mbox{ex}}$. As we shall see below,
the picture of the moving singularity get too smooth and hardly yields to
observation.

However, the situation sharply varies if not the s--wave but p-- or higher
waves dominate in the amplitude of secondary interaction. The shape of the
cusp singularity changes and it ceases to dominate at small $E_{\mbox{ex}}$.
As will be shown below, the situation with observation of the moving
triangle singularity is essentially improved and we can see very clear and
unambiguous picture. Later on it will be illustrated on graphic examples.
Yet at first we will develop a formalism for Fig.1 graph calculation with
arbitrary nuclear form factor in A$\to 1+3$ vertex and account of nonzero
orbital momenta in the lower vertex.

Except the quantities $M$, $E_{\mbox{ex}}$ and ${\bf q}= {\bf p_x} -
{\bf p_y}$ (all momenta in lab. system), introduced earlier, we will also
use the nuclear binding energy ${\varepsilon}=m_1 + m_3 - m_A$ and
the characteristic momentum $\kappa$ in the nuclear vertex $A \to 1+3$:
${\kappa}^2= 2m_{13}{\varepsilon}$,
where $m_{13}$ is the reduced mass of particles 1 and 3.
We will usually consider nuclear reactions and take the particles A, 1 and
3 to be nonrelativistic. It is convenient to use the dimensionless
variables $\xi$ and $\lambda$ [9,7], associated with $M$ and $q$:
\begin{eqnarray}
\xi  \equiv  \xi (M,q) & = & \frac{m_A}{2 m_3 \varepsilon} \, 
\left\{ \frac{m_1q^2}{M^2+q^2} +
\frac{M^2+m_1^2-{\mu}^2}{\sqrt{M^2+q^2}} -2m_1 \right\} \, , \\
\lambda  \equiv  \lambda (M,q) & = & \frac{m_1^2}{M^2+q^2} \,
\frac{q^2}{{\kappa}^2}  \, .
\end{eqnarray}
These expressions are valid in general case of relativistic particle 2.
In the nonrelativistic case $\xi$ is expresses only through $M$:
\begin{equation}
\xi \equiv {\xi}(M)  = \frac{m_2}{m_3} \,\frac{m_A}{M } \,
\frac{M -m_1-m_2}{\varepsilon} \, .
\end{equation}
In terms of these variables the triangle graph has a square root
branching point at  $\xi =0$ and logarithmic singularities at
$\xi = \lambda -1 \pm i \sqrt{\lambda} \, .$

Let us go to the estimation of Fig.1 graph for the case of arbitrary
upper left (nuclear) and lower (secondary interaction) vertices. We will
take the upper right (primary interaction) vertex to be constant,
equivalent to factoring out its value in the triangle integral at some
point near the maximum of the integrand. Usually this means that we
factor out the amplitude of the process $ x+3 \to y+2$ at the proper
momentum transfer and at the energy corresponding to the interaction of
particle $x$ with particle 3 at rest. The amplitude of the graph has the
form of a four-fold integral over the 4-momentum $p$ of particle 3 [10].
The integration over the fourth component reduces to determining the
residue in the pole of the propagator of particle 1. (Then neglected terms
are of the order $\sqrt{{\varepsilon}/m}$ or ${\varepsilon}/{\mu}$ where
$m$ and $\mu$ are nucleon and pion masses.) So in reality we have
three--fold integral over ${\bf p}$, containing propagators of particles
2 and 3 and two nontrivial vertices. The nuclear vertex $A \to 1+3$
together with the propagator of particle 3 corresponds to a nuclear wave
function $ {\Psi}_l(p)Y_{lm}({\hat {\bf p}})$.
The lower vertex can be represented as
\begin{equation}
{\varphi}_L(k)Y_{LM}^{*}({\hat {\bf k}})
\end{equation}
where ${\bf k}$ is the relative momentum of particles 1 and 2
\begin{equation}
{\bf k}={\bf p} + {\bf Q}, \qquad  {\bf Q}=\frac{m_1}{m_1+m_2} {\bf q} \, .
\end{equation}
Here $l$ and $L$ are the orbital momenta of the nuclear vertex and the
secondary interaction vertex. As for the latter, we deal here only with
nontrivial part (4)
of its amplitude which contains the dependence on ${\bf k}$. The most
interesting phenomena appear in the $M$ region not far from the threshold
of the channel (1+2), i.e. for $m_1+m_2$ not far from $M$, where the
dependence on ${\bf k}$ is most important.  Actually all vertices
also contain spin parts (Clebsch - Jordan coefficients due, for example, to
spin--orbital interaction), but these do not influence the
integration procedure.

The propagator of particle 2 can be transformed to the form
\begin{equation}
2m_2E_2- {\bf p}_2^2 +i{\eta}
=-\frac{m_1+m_2}{m_1} \{ ({\bf p}+{\bf Q})^2 -2m_{12}E_{ex} - i{\eta}
\} \, .
\end{equation}
Note that $ Q^2={\lambda}{\kappa}^2,\quad 2m_{12}E_{ex}= {\xi}{\kappa}^2
\, . $ The propagator (6) contains the same vector combination as in
eq.(5) that is the relative momentum of particles 1 and 2. This is
important for our formal transformations. This statement is valid
also in the case of a relativistic particle 2 (see ref. [9]). The
expression for the amplitude of the triangle graph of Fig.1 is
proportional to
\begin{equation}
I_{\triangle} = \int d{\bf p} \frac{{\Psi}_l(p)Y_{lm}({\hat {\bf p}})
{\varphi}_L(|{\bf p}+{\bf Q}|)Y_{LM}^{*}({\bf p}+{\bf Q})}
{({\bf p}+{\bf Q})^2-2m_{12}E_{ex}-i{\eta}} \, .
\end{equation}
Here $E_{ex}$ can take positive as well as negative values.

Let us transform eq.(7) to coordinate space and use nuclear wave
function in coordinate space ${\psi}_l(r)Y_{lm}({\hat {\bf r}})$:
\begin{equation}
{\Psi}_l(p)Y_{lm}({\hat {\bf p}}) = \int {\psi}_l(r')Y_{lm}
({\hat {\bf r}}')e^{i{\bf p}{\bf r}'} d{\bf r}' \, .
\end{equation}
If the lower vertex were constant we would use one of the
following relations to transform the denominator of eq.(7):
\begin{equation}
\frac{1}{{\bf k}^2-{\alpha}^2 -i{\eta}} = \frac{1}{4{\pi}}
\int \frac{e^{i{\bf k}{\bf r}+i{\alpha}r}}{r}\, d{\bf r} \, , \qquad
\frac{1}{{\bf k}^2+{\beta}^2}  = \frac{1}{4{\pi}}
\int \frac{e^{i{\bf k}{\bf r}-{\beta}r}}{r}\, d{\bf r}\, .
\end{equation}
Then simple calculation would lead to the result  (for example if $E_{ex}<0$)
\begin{equation}
I_{\triangle}=(2{\pi})^3 i^l Y_{lm}(-{\bf Q}) \, \int_0^{\infty}dr \, r j_l(\sqrt{\lambda}
{\kappa} r) exp(-\sqrt{-{\xi}} {\kappa} r) {\psi}_l(r) \, .
\end{equation}
For the case $E_{ex}>0$ the quantity $-i \sqrt{\xi}$ must be substituted
for  $\sqrt{-{\xi}}$.

For an arbitrary lower vertex we will apply a transformation to
coordinate space which includes a new function $f_L(r)$ which will be
described later:
\begin{equation}
\frac{{\varphi}_L(k)Y_{LM}^{*}({\hat {\bf k}})}{k^2+{\beta}^2 } =
\int Y_{LM}^{*}({\hat {\bf r}})f_L(r)e^{-i{\bf k}{\bf r}} d{\bf r} \, .
\end{equation}
Then straightforward but tedious calculations lead to the result
\begin{eqnarray}
&& I_{\triangle}=(2{\pi})^3\sqrt{4{\pi}(2l+1)}(-1)^l\sum_{T,t}
i^T C_{l0L0}^{T0}C_{lmTt}^{LM}
Y^{*}_{Tt}(-{\hat {\bf Q}}) \\  \nonumber
&& \qquad \cdot \int_0^{\infty} dr\, r^2 j_T(Qr){\psi}_l(r)f_L(r)
\, 
\end{eqnarray}
where $j_T$ is a spherical Bessel function.
The nontrivial part of $I_{\triangle}$, which will be denoted as ${\cal M}$
\begin{equation}
{\cal M} = \int\limits_0^{\infty} dr\cdot r^2 j_T(Qr)
{\psi}_l(r)f_L(r) \, ,
\end{equation}
contains one-fold integral with radial part of nuclear wave function and
function $f_L(r)$ associated with the lower vertex of the triangle graph of
Fig.1. Here $T$, $l$ and $L$ must obey the triangle rule. The
relative weights of terms with different $T$ in the final results
for the cross section, polarizations etc. will depend not only on factors
in eq.(12) but also on the spin structure of all vertices in the triangle
graph.

Let us turn to eq.(11). Inverse transformation gives
\begin{equation}
f_L(r)=\frac{i^L}{2{\pi}^2}\int\limits_0^{\infty}
\frac{{\varphi}_L(k)j_L(kr)} {k^2+{\beta}^2} k^2 \, dk \, .
\end{equation}
Various choices of this function are described in ref. [9]. Here we consider
only one kind chosen such as to give a correct behavior of the amplitude
for the case of arbitrary orbital momenta in the near--threshold region:
\begin{equation}
{\varphi}_L(k)=C\, k^L.
\end{equation}
Then for example
\begin{equation}
f_0(r)=\frac{C}{4\pi} \frac{e^{-{\beta}r}}{r}=-\frac{C{\beta}}{4 \pi} h_0^{(1)}(i{\beta}r).
\end{equation}
In the general case,  a minimum account of the centrifugal barrier
requires
\begin{equation}
f_L(r)= (-1)^{(L+1)} C \frac{{\beta}^{L+1}}{4 \pi} h_L^{(1)} (i{\beta}r) \, ,
\end{equation}
where $h_L^{(1)}$ is the spherical Hankel function of the first class.
Expressions (11), (14), (16), (17) correspond to ${\xi}<0$ when
${\beta}= \sqrt{-{\xi}} {\kappa}$. For the case ${\xi}>0$ the quantity
$\beta$ in these expressions must be
changed into $-i{\gamma}$ where ${\gamma}=\sqrt{\xi} {\kappa}$.

The formula (13) enables us to carry out numerical estimations in a simple
way. Here we will show the results of concrete calculations for several
cases with s-- and p-- orbital momenta. For s-wave nuclear wave functions we
used the deuteron function in Paris potential as well as the wave function
in Gauss parametrization ${\psi}(r) \sim exp(-p_0^2 r^2/2)$.
For the p-wave we used a p-wave Gauss function
${\psi}(r) \sim r exp(-p_0^2 r^2/2)$
as well as a ``quasi-Hulten'' p-wave function of the form
\begin{equation}
{\psi }(r) \sim \left(\frac{1}{{\kappa}r} + \frac{1}{{\kappa}^2r^2} \right)
\left( e^{-{\kappa}r}
-3e^{-({\kappa}+{\rho})r}+ 3e^{-({\kappa}+2{\rho})r}- e^{-({\kappa}+
3{\rho})r} \right) \,
\end{equation}
which has correct asymptotic behavior at $r \to 0$ and $r \to \infty$.

Fig.2 shows the results for $|{\cal M}|^2$ for the case of the deuteron
with Paris wave function: a) for s-wave lower vertex and b) for p-wave
lower vertex. In the latter case
there is now a distinctive and pronounced systematic pattern in  which
the maximum of the distribution moves to larger values of
$\xi$ with increasing $\lambda$. For s-wave Gauss function the pattern
are close to the deuteron case. Fig.3 illustrates the corresponding effect
for the case of p-wave Gauss nuclear function with $p_0=0.75 {\kappa}$
(it corresponds to $^{12}$C).
Again the picture is more distinct for the case of p-wave lower
vertex. The p-wave ``quasi--Hulten'' function (18) gives similar results.
These calculations were done with the amplitude of lower vertex of the
Fig.1 graph in the form (22). Introduction of
cut--off factors (see ref. [9]) does not lead to the qualitative changes.
We see that the cases with the p--wave lower vertex open the possibility
to observe the picture of the moving triangle singularity experimentally.
Up to now this picture  has not been exploited for the identification of
the triangle mechanism.

In conclusion it can be said that
we have first derived the simple and transparent formulas (12)-(14) for the
triangle graph with arbitrary orbital momenta in the nuclear vertex and in
the vertex of secondary interaction. The amplitude is represented in the form
of one-fold integral in coordinate space. It is well suitable for numerical
calculations due to the exponential decrease of the nuclear wave function.

We then have pointed out that nonzero orbital momenta in the vertices have an
important effect on the shape of experimental distributions. This idea is
supported by the graphical results. The corresponding
singularity manifests itself much more distinctly and sharply than for the
case with $L=0$. The experimental demonstration of this effect requires
favourable situations. This is not easy, since the most interesting region
is located close to the threshold for particles $1+2$, where the p--wave
amplitude is normally suppressed by the barrier.
One interesting example is ${\bar N}N$--interaction in the final state for
which the p-wave is greatly enhanced at low energies and can be separated
by the choice of  particular channels [11]. A second example is the
${\pi}N$--interaction which is dominated by p-wave part from rather small
energies.

The author is indebted to Prof. O.D.Dalkarov, T.E.O.Ericson and I.S.Shapiro
for discussions. He also appreciate the hospitality of The Svedberg
Laboratory of the Uppsala University where a part of this investigation
was done.

\newpage
\begin{figure}
\begin{picture}(250,250)
\put(170,15){\epsfig{file=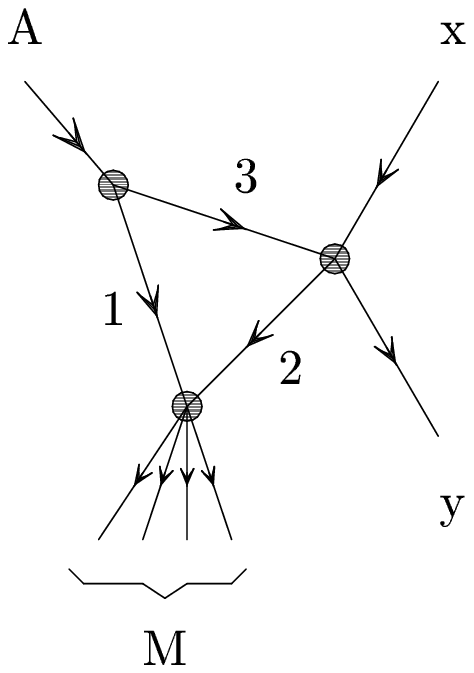}}
\end{picture}
\caption{Typical triangle graph.}
\end{figure}

\begin{figure}
\begin{picture}(400,500)
\put(30,15){\epsfig{file=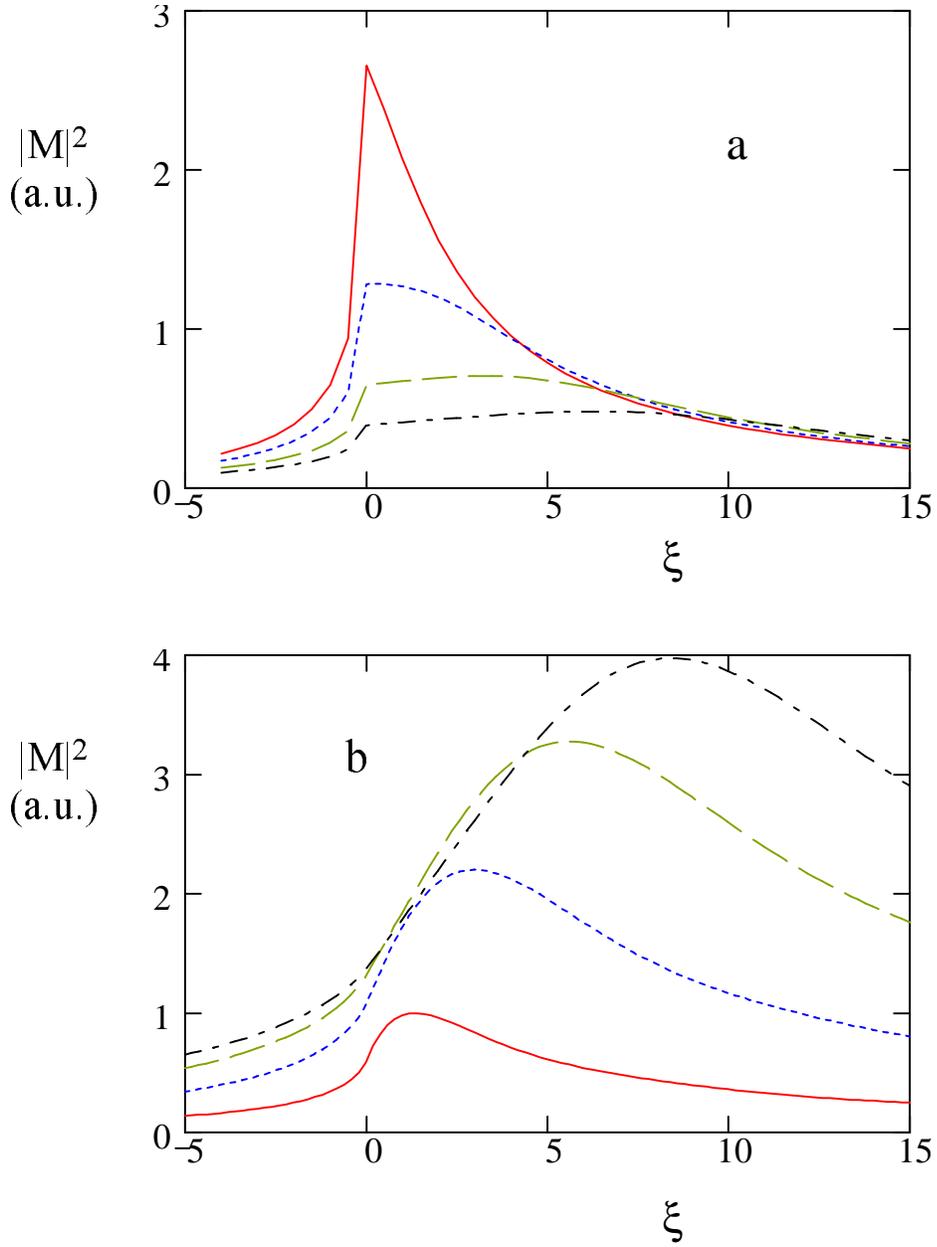}}
\end{picture}
\caption{$|{\cal M}|^2$ as a function of ${\xi}$ for ${\lambda}=1$
(solid line), 3 (dotted line), 6 (dashed line), 9 (dash-dotted line)
in the case of the deuteron with Paris wave
function: a) for s-wave lower vertex, b) for p-wave lower vertex.}
\end{figure}

\begin{figure}
\begin{picture}(400,500)
\put(30,15){\epsfig{file=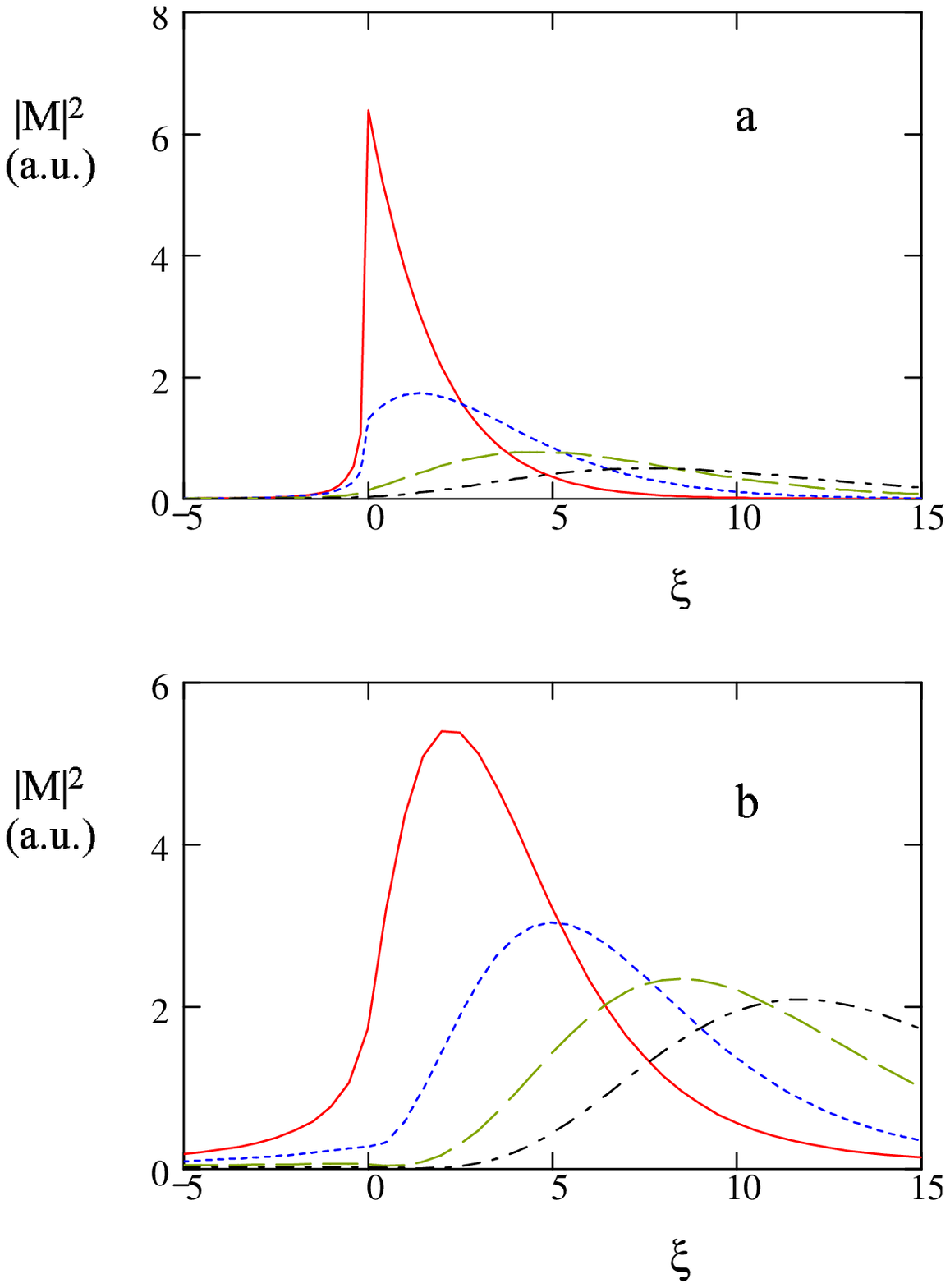}}
\end{picture}
\caption{The same as in Fig.2 for p-wave nuclear function in
the Gauss parametrization: a) for s-wave lower vertex, b) for p-wave
lower vertex.}
\end{figure}

\end{document}